\documentclass[a4paper,fleqn,usenatbib]{mnras}

\usepackage{mathtools}  
\usepackage{amssymb}
\usepackage[T1]{fontenc}
\usepackage{ae,aecompl}
\usepackage{adjustbox}

\usepackage{graphicx}	
\usepackage{amsmath}	
\usepackage{amssymb}	
\usepackage{natbib} 
\usepackage{lipsum}

\title[Polarimetry through multi-line observations III]{Chromospheric polarimetry through multi-line observations of the 850~nm spectral region III: Chromospheric jets driven by twisted magnetic fields}

\author[C. Quintero Noda et al.]{C. Quintero Noda,$^{1,2,3}$\thanks{E-mail: c.q.noda@astro.uio.no}
H. Iijima,$^{4}$
Y. Katsukawa,$^{5}$
T. Shimizu,$^{1}$
M. Carlsson,$^{2,3}$
\newauthor
J. de la Cruz Rodr\'{i}guez,$^{6}$
B. Ruiz Cobo,$^{7,8}$
D. Orozco Su\'arez,$^{9}$
T. Oba,$^{1}$
\newauthor
T. Anan,$^{10}$
M. Kubo,$^{5}$
Y. Kawabata,$^{1,11}$
K. Ichimoto,$^{5,12}$
Y. Suematsu$^{5}$ 
\\
$^{1}$Institute of Space and Astronautical Science, Japan Aerospace Exploration Agency, Sagamihara, Kanagawa 252-5210, Japan\\
$^{2}$Rosseland Centre for Solar Physics, University of Oslo, P.O. Box 1029 Blindern, N-0315 Oslo, Norway\\
$^{3}$Institute of Theoretical Astrophysics, University of Oslo, P.O. Box 1029 Blindern, N-0315 Oslo, Norway\\
$^{4}$Institute for Space-Earth Environmental Research, Nagoya University, Furocho, Chikusa-ku, Nagoya, Aichi 464-8601, Japan\\
$^{5}$National Astronomical Observatory of Japan, 2-21-1 Osawa, Mitaka, Tokyo 181-8588, Japan\\
$^{6}$Institute for Solar Physics, Dept. of Astronomy, Stockholm University, Albanova University Center, SE-10691 Stockholm, Sweden\\
$^{7}$Instituto de Astrof\'isica de Canarias, E-38200, La Laguna, Tenerife, Spain.\\
$^{8}$Departamento de Astrof\'isica, Univ. de La Laguna, La Laguna, Tenerife, E-38205, Spain\\
$^{9}$Instituto de Astrof\'isica de Andaluc\'ia (CSIC), Glorieta de la Astronom\'ia, 18008 Granada, Spain\\
$^{10}$National Solar Observatory, 22 Ohi'a Ku, Makawao, HI 96768, USA\\
$^{11}$Department of Earth and Planetary Science, The University of Tokyo, 7-3-1 Hongo, Bunkyo-ku, Tokyo 113-0033, Japan\\
$^{12}$Kwasan and Hida Observatories, Kyoto University, Kurabashira Kamitakara-cho, Takayama-city, 506-1314 Gifu, Japan\\
}
\date{Accepted XXX. Received YYY; in original form ZZZ}

\pubyear{2018}

\begin{document}
\label{firstpage}
\pagerange{\pageref{firstpage}--\pageref{lastpage}}
\maketitle

\begin{abstract}
We investigate the diagnostic potential of the spectral lines at 850~nm for understanding the magnetism of the lower atmosphere. For that purpose, we use a newly developed 3D simulation of a chromospheric jet to check the sensitivity of the spectral lines to this phenomenon as well as our ability to infer the atmospheric information through spectropolarimetric inversions of noisy synthetic data. We start comparing the benefits of inverting the entire spectrum at 850~nm versus only the Ca~{\sc ii}~8542~\AA \ spectral line. We found a better match of the input atmosphere for the former case, mainly at lower heights. However, the results at higher layers were not accurate. After several tests, we determined that we need to weight more the chromospheric lines than the photospheric ones in the computation of the goodness of the fit. The new inversion configuration allows us to obtain better fits and consequently more accurate physical parameters. Therefore, to extract the most from multi-line inversions, a proper set of weights needs to be estimated. Besides that, we conclude again that the lines at 850~nm, or a similar arrangement with Ca~{\sc ii}~8542~\AA \ plus Zeeman sensitive photospheric lines, poses the best observing configuration for examining the thermal and magnetic properties of the lower solar atmosphere.
\end{abstract}


\begin{keywords}
Sun: chromosphere -- Sun: magnetic fields -- techniques: polarimetric
\end{keywords}



\section{Introduction}

\begin{figure*}
\begin{center} 
 \includegraphics[trim=0 0 0 0,width=17.0cm]{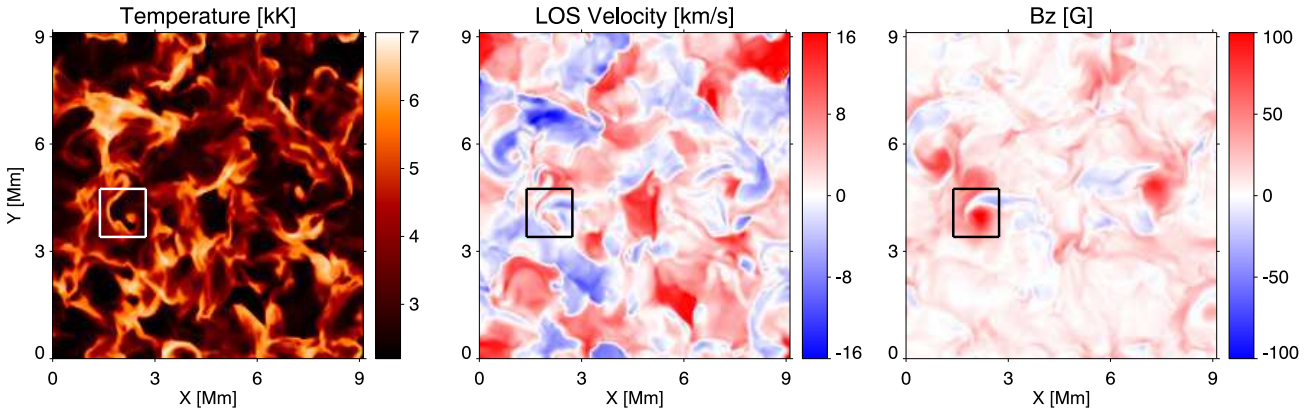}
 \vspace{-0.1cm}
 \caption{From left to right, spatial distribution of temperature, line-of-sight velocity, and the vertical component of the magnetic field at $\log \tau=-5$, respectively. Negative LOS velocities correspond to material moving upwards with respect to the solar surface.  The highlighted location is examined later.}
 \label{fov}
 \end{center}
\end{figure*}

This is the third study of a series of publications investigating the properties of the spectral lines belonging to the 850~nm window. In the first paper, we examined their diagnostic potential for inferring the atmospheric parameters at photospheric and chromospheric layers with semi-empirical models and more complex atmospheres as the 3D enhanced network simulation \citep{Carlsson2016} generated with the {\sc bifrost} code \citep{Gudiksen2011}. We concluded that, although the Ca~{\sc ii} 8542~\AA \ is already a very capable spectral line \citep[e.g.,][]{Cauzzi2008}, we can improve its diagnostic capabilities observing the neighbouring spectrum. This is because there are nearby lines that enhance the sensitivity to the atmospheric parameters at lower heights as well as at upper atmospheric layers \citep[see, for instance, the comparative work by][]{SocasNavarro2000b}. 

In the second study of this series, we examined the two-dimensional simulation of a magnetic flux tube presented in \cite{Kato2016}. In this case, the atmosphere inside the magnetic concentration periodically changes due to the so-called magnetic pumping process. We determined that this variation of the atmospheric conditions leaves a characteristic imprint in the Stokes profiles of photospheric and chromospheric spectral lines through large Dopplershifts and intensity fluctuations \citep{QuinteroNoda2017b}. Also, we concluded both publications mentioning that we need to perform additional studies to understand better our capabilities for inferring the atmospheric information from spectropolarimetric observations. In particular, to assess the accuracy of the inferred atmospheric parameters through non-local thermodynamic equilibrium (NLTE) inversions of multiple spectral lines with a different height of formation \cite[see, for instance, the recent studies by][]{daSilvaSantos2018,Leenaarts2018}. Our target is to pave the road for when chromospheric polarimetric observations are routinely performed in the future. Something that will happen soon when ground-based telescopes such as DKIST \citep{Keil2011} or EST \citep{Collados2013}, as well as balloon missions such as the Sunrise solar balloon-borne observatory \citep{Barthol2011,Berkefeld2011,Gandorfer2011}, start performing observations. 

In this publication, we focus on the MHD simulation of a chromospheric jet with complex plasma flows and twisted magnetic field lines by \cite{Iijima2016,Iijima2017}. In this simulation, the thermal convection near the solar surface excites various MHD waves and generates chromospheric jets rooted in magnetic field concentrations. Also, the magnetic field lines that define the jet are entangled at chromospheric heights, which helps the chromospheric jet to be driven by the Lorentz force. 

The primary target of this study is to examine the diagnostic potential of the 850~nm spectral lines for inferring the physical information of the chromospheric jet through numerical NLTE inversions of noisy synthetic spectra. To this purpose, we analyse one of the dynamic events of the simulation where large gradients dominate the atmospheric parameters, and we propose an inversion set-up that optimally works when performing simultaneous inversions of the photospheric and chromospheric spectral lines belonging to the 850~nm window.

\section{Simulations and methodology}

\cite{Iijima2017}  developed a 3D radiation MHD simulation with the numerical code RAdiation Magnetohydrodynamics Extensive Numerical Solver \citep[{\textsc{ramens}},][]{Iijima2015} that extends from the upper convection zone to the lower corona, i.e. from -2 to 14~Mm above the solar surface. The setup of this simulation is explained in \cite{Iijima2017} although we describe below some of its properties for the sake of clarity. 

The horizontal domain of the simulation covers $9\times9$~Mm$^{2}$ with a uniform grid size of 41.7~km, i.e. a spatial resolution of around 80~km.  However, the effective spatial resolution is slightly larger than twice the grid size (typically 3-6 grid zones, or 120-240 km in this case). This reduction of the effective resolution is caused by the numerical diffusion of the MHD scheme. Therefore, the effective spatial resolution is comparable to the values that can be achieved with current 1~m or larger telescopes (i.e. Sunrise, SST \citep{Scharmer2003}, GST \citep{Goode2003}, or Gregor \citep{Schmidt2012}) and indeed will be attained by future large 4~m telescopes as DKIST and EST. As an illustrative example, SST can achieve, observing the 850~nm window, a spatial resolution of 0.2~arc sec at diffraction limit, i.e. around 150~km.

The vertical domain comprises 16~Mm with a constant step size of 29.6~km. Horizontal boundary conditions are periodic while the top and bottom boundaries are open to flows. In addition, the computation is done in multiple stages although we only analyse the simulation results corresponding to the last 30 minutes of the run. Figure~\ref{fov} depicts an example of the whole simulation horizontal domain at $\log \tau=-5$, where $\tau$ is the optical depth at 500~nm. The temperature panel shows a complex scenario covered by ubiquitous thin and hot structures surrounded by cool areas. In the case of the line-of-sight (LOS) velocity panel, we can see a dynamic pattern with upflows and downflows of more than 16 km/s in a few locations with, in general, the former associated with hot areas. If we move to the longitudinal component of the magnetic field, we can detect two main concentrations of the same polarity with opposite sign magnetic fields in between them. In both cases, the longitudinal component of the magnetic field shows values close to 100~G while it is weaker outside those concentrations being lower than 10~G. 

We focus on the region enclosed by a box (see Figure~\ref{fov}) that is centred on the chromospheric jet studied in \cite{Iijima2017}. This feature has a maximum height of 10~Mm, and its core is hotter than its surroundings at chromospheric layers. Also, it shows an intricate magnetic structure with entangled and twisted magnetic field lines (see Figure~8 of the mentioned work). 

\subsection{Synthesis of the Stokes profiles}

The method used in the present study is similar to the one employed in \cite{QuinteroNoda2017a,QuinteroNoda2017b}. We generate the full Stokes vector for the 850~nm window shown in Figure~1 of the first publication of this series. Inside this spectral window, there are several lines of interest, with the most important ones being the photospheric Fe~{\sc i} transitions at 8468 and 8514~\AA. They show high sensitivity to the magnetic field (comparable to, e.g., Fe~{\sc i} 5250.2 or 6302.5~\AA) at lower layers. Moreover, we can also find inside the 850~nm spectral region two chromospheric lines from the infrared Ca~{\sc ii} triplet at 8498 and 8542~\AA \ that are sensitive to the atmospheric parameters up to the middle chromosphere, around 1000~km above the visible surface.

We perform synthesis of these spectral lines with the {\sc nicole} code \citep{SocasNavarro2000,SocasNavarro2015}. This is done with column-by-column forward modelling, i.e. each column is treated independently, and the non-local thermodynamic equilibrium atomic populations are solved for the Ca~{\sc ii} line transitions assuming a plane-parallel atmosphere. This approximation is appropriate for the Ca~{\sc ii} lines where horizontal scattering does not represent a dominant contribution \citep{Leenaarts2009,delaCruzRodriguez2012}. Also, the code works under the field-free approximation \citep{Rees1969}, i.e. the statistical equilibrium equations are solved neglecting the presence of a magnetic field \citep[see also][for more details]{Bruls1996,TrujilloBueno1996}. 

The spectral sampling used is the same as well, i.e. $\Delta\lambda=40$~m\AA. Therefore, the full spectrum of interest would fit in a hypothetical sensor of 2000~pixels (for instance, Katsukawa et al. in preparation). The simulation's vertical domain is shortened to $z=[-650,2500]$~km because the spectral lines of interest form within that range of heights and this allows us to reduce the computational time. We assume that we are looking at the disc centre, i.e. $\mu=1$ (where $\mu=\cos\theta$, and $\theta$ is the angle of the ray with respect to the normal of the atmosphere). We deactivate the default ``velocity free" option because we believe it would produce less accurate results in this simulation where strong LOS velocity gradients are dominant. We switch on the keyword ``optimise grid" that interpolates the initial atmosphere to a different grid of points in height. The resulting atmosphere is the one used in any plot of this manuscript, including that of Figure~\ref{fov}. We start the initial guess for the atomic populations assuming LTE populations using the {\sc multi} \citep{Carlsson1986} approach (see {\sc nicole}'s manual for more information). Also, line broadening owing to collisions by neutral hydrogen atoms is applied to all the spectral lines following the theory of \cite{Barklem1998}.

To simulate the effect of a general spectral point spread function, we degrade the spectra employing a Gaussian profile with a full width at half maximum of 1.0~km/s, and we use the original spatial sampling, i.e. no spatial degradation. We also simulate the presence of noise in observations adopting similar conditions to those expected for the Sunrise Chromospheric Infrared spectro-Polarimeter (SCIP, Katsukawa et al. in preparation) instrument under development for the third flight of the Sunrise balloon. In this regard, the nominal observing mode of SCIP integrates longer than 10 s per slit position and aims to achieve a noise level for the polarisation Stokes profiles of the order of $\pm3\times10^{-4}$ of $I_c$. Thus, this is the noise value we assume for Stokes ($Q$, $U$, $V$). In the case of Stokes $I$, there are always systematic errors in the post-processing of the signals that prevent achieving so low noise levels. Therefore, we tentatively choose a random noise with a standard deviation of $\pm1\times10^{-2}$ of $I_c$. Those noise signals are different for each spectral point, each Stokes profile, and each spatial pixel.
 
\begin{figure}
\begin{center} 
 \includegraphics[trim=0 0 0 0,width=7.8cm]{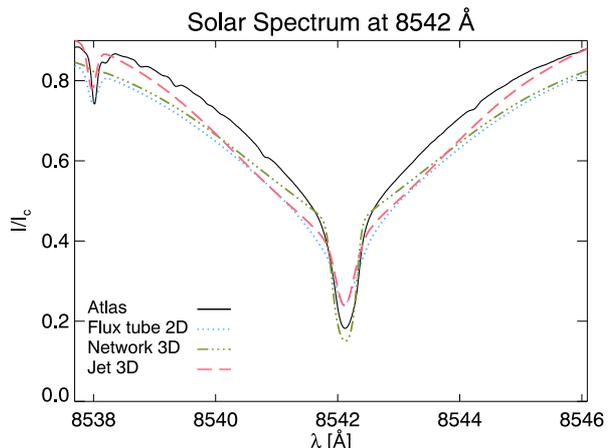}
 \vspace{-0.2cm}
 \caption{Comparison between the solar atlas (black) and the spatially averaged intensity profile for different simulations (colour).}
 \label{Atlas_compare}
 \end{center}
\end{figure}

\begin{figure*}
\begin{center} 
 \includegraphics[trim=0 0 0 0,width=17.5cm]{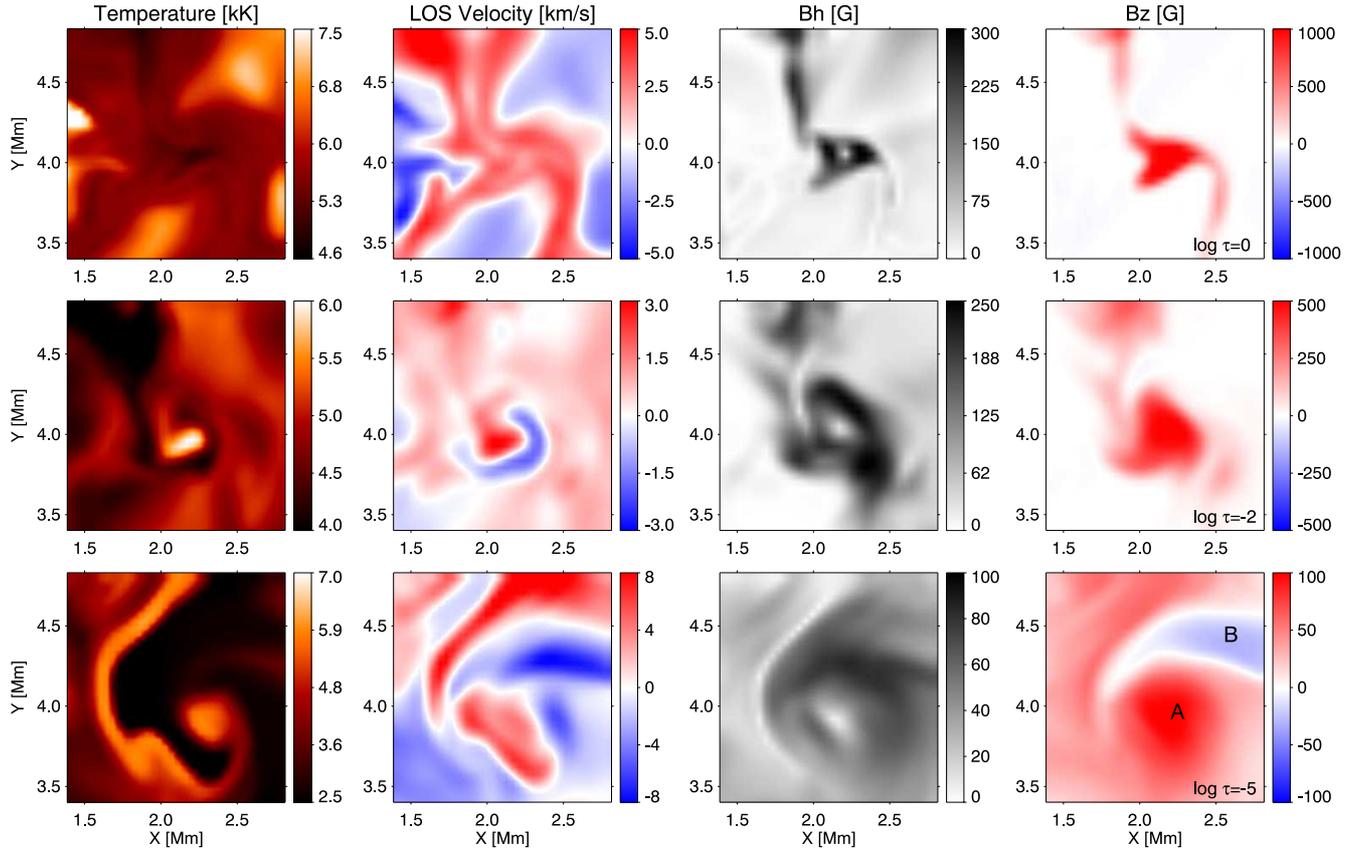}
 \vspace{-0.2cm}
 \caption{Each panel shows the spatial distribution, from left to right column, of temperature, LOS velocity, horizontal and vertical components of the magnetic field at the optical depths $\log \tau=[0,-2,-5]$. Negative LOS velocities designate material moving upwards with respect to the solar surface. The field-of-view corresponds to the region highlighted in Figure~\ref{fov} while Roman characters indicate the main longitudinal components of the chromospheric jet at upper layers.}
 \label{Context}
 \end{center}
\end{figure*}

\subsection{Line core width}

\cite{Leenaarts2009} explained that current numerical simulations do not contain sufficient heating and small-scale motions to match the observed intensities and widths of chromospheric lines. Moreover, if the line core intensity profile is narrower and deeper than expected, this could induce unrealistic polarisations signals \citep{delaCruzRodriguez2012}. Therefore, we follow the steps of the first two publications of this series examining first the intensity profile spatially averaged over the whole simulation box presented in Figure~\ref{fov}, synthesised using a null microturbulence value. The results are depicted by the dashed line in Figure~\ref{Atlas_compare}, while solid corresponds to the solar atlas \citep{Delbouille1973}, dashed-dotted line displays the results from the spatially averaged profile of snapshot 385 of the enhanced network simulation \citep{Carlsson2016} using a microturbulence of 3~km/s, and dots represent the averaged profile computed using the 2D flux sheath simulation \citep{Kato2016} and a null microturbulence. 
 
The intensity profile produced by the chromospheric jet simulation (dashed) is broad, showing a line core width at full width half maximum of 423~m\AA, close to the value displayed by the solar atlas, i.e. 574 m\AA, and the results of additional studies \citep[for instance, ][ obtained a line core width ranging between 450-550~m\AA]{Cauzzi2009}. Therefore, in this case, we do not need to introduce an additional microturbulence contribution.  Finally, although the width of the line is sufficiently broad to avoid using an artificial microturbulence, we can still detect intensity differences in the outer wings of the line, e.g. at 8540 and 8544~\AA. We guess that this could be caused because none of those simulations aims to represent strictly quiet Sun conditions.

\section{Chromospheric jet}

\subsection{Atmospheric parameters}

We are interested in the enclosed region highlighted in Figure~\ref{fov}. In particular, we examine the spatial distribution of different atmospheric parameters at selected heights in the atmosphere. We focus on the temperature, LOS velocity, horizontal ($B_{\rm h}=\sqrt{B_{x}^2+B_{y}^2}$) and vertical ($B_z$) components of the magnetic field (see columns in Figure~\ref{Context}) at three selected optical depths, i.e. $\log \tau=[0,-2,-5]$. Those atmospheric layers correspond to the reference continuum wavelength, the height of formation of commonly observed  photospheric lines as Fe~{\sc i} 6301.5~\AA \ \citep{Khomenko2007}, and approximately the height of formation of the Ca~{\sc ii} 8542~\AA \ line \citep[][]{Cauzzi2008}, respectively.

\begin{figure*}
\begin{center} 
 \includegraphics[trim=0 0 0 0,width=16.5cm]{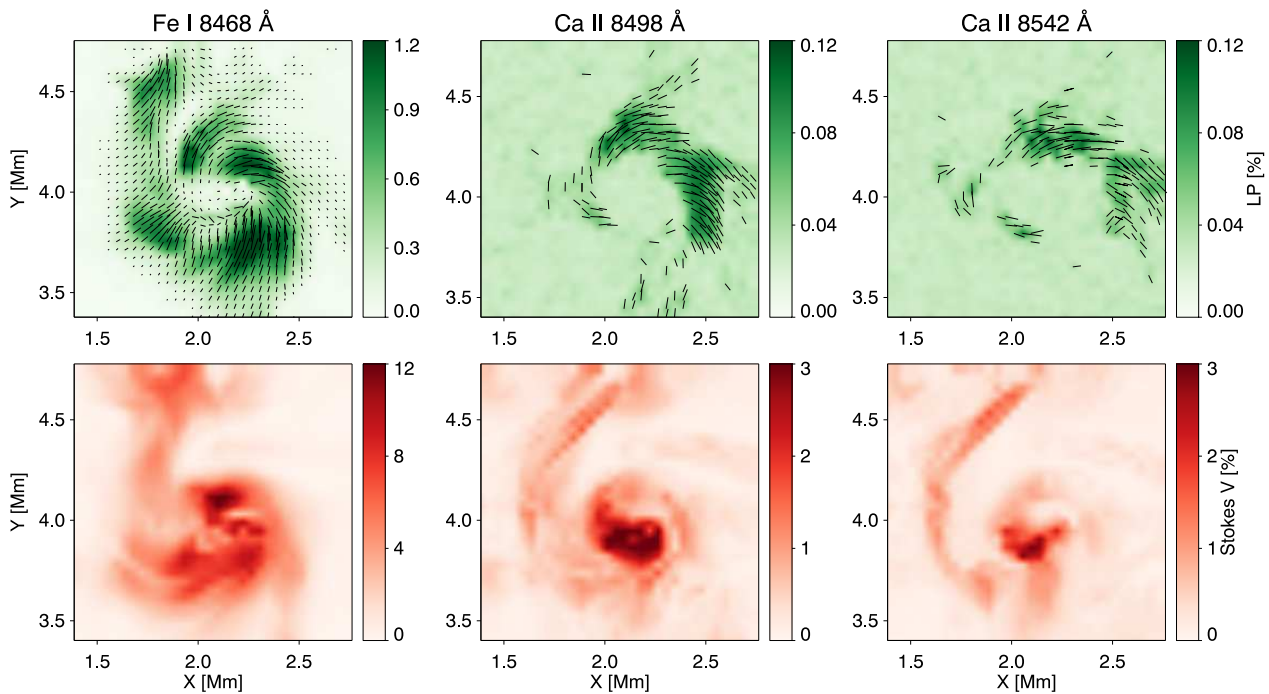}
 \vspace{-0.2cm}
 \caption{Panels show the spatial distribution of maximum linear (top) and circular (bottom) polarisation signals for selected spectral lines. From left to right, Fe~{\sc i} 8468~\AA, and the two Ca~{\sc ii} infrared lines at 8498 and 8542~\AA, respectively. We display the inferred average magnetic field azimuth with headless arrows in the top panel for those pixels with an amplitude larger than $3\times10^{-4}$ of I$_c$. The magnitude of the linear polarisation signals gives the length of each arrow.}
 \label{Azi}
 \end{center}
\end{figure*}

Starting with the spatial distribution of the gas temperature (leftmost column), we find that the granulation is well defined in the selected field-of-view, being the location of the jet, i.e. in $x$ and $y$ coordinates equal to (2.2,4.0)~Mm, defined by a cool intergranular region. Moving on to upper layers (middle row), we detect areas with enhanced temperature, with a localised hot point at the roots of the chromospheric jet, but also a cool patch at around (1.6,4.6)~Mm. Higher in the atmosphere (bottom row), there is a thin hot ``thread'' surrounding the core of the jet. Additional cool areas can be found at different spatial locations, for instance, close to the centre of the jet at (1.9,4.2)~Mm. 

Concerning the LOS velocity, the granulation pattern is also discernible. At $\log \tau=-2$ the LOS velocity is characterised by low amplitude velocities, less than 2~km/s in absolute value, that is in general represented by material moving downwards (see red patches). However, we can detect a stronger downflow of roughly 4~km/s surrounded by upflowing material at (2.2,3.8)~Mm, that corresponds to the jet. This also takes place at higher layers, bottom panel, where a similar pattern now occupies broader areas. 

In the case of the horizontal component of the magnetic field (third column), it is concentrated in a narrow region in the lower photosphere with values larger than 300~G at some points, e.g. (2.3,4.0)~Mm. This region expands occupying wider areas at $\log \tau=-2$, and the field strength drops to values in between $250-300$~G. In the chromosphere (bottom), the spatial distribution shows almost null values in the same spatial location of the core of the jet, i.e. (2.2,4.3)~Mm, with a field strength of the order of  tens of Gauss.

The longitudinal magnetic field (rightmost column), displays a similar behaviour as $B_{\rm h}$, i.e. localised concentrations at lower layers that expand with height. In addition, in the case of the jet, a single polarity component dominates lower layers. At higher heights, i.e. $\log \tau=-5$, there are two main components of opposite polarity, a positive one that corresponds to the core of the jet at around (2.2,4.0)~Mm (label A), and a negative one located near the jet (label B) that extends towards the right side beyond the selected field-of-view. Also, the interface between those two opposite polarity regions is comprised by almost null $B_z$ values, co-spatial with areas of larger $B_{\rm h}$ (see the bottom panel in the third column).  In this regard, if we look at Figure 8 of \cite{Iijima2017} the A region is associated to the central spine of the jet while the B labelled area corresponds to the almost isolated magnetic field lines that, after twisting around the jet, extend beyond it (see sky blue lines in the cited figure around $Z$=1~Mm).

\subsection{Spectral features}

We plan to study in this section the properties of the Stokes profiles that can reveal information about the jet. In particular, the magnetic field azimuth and its longitudinal component. The former one can be inferred using the ratio of the synthetic Stokes $Q$ and $U$ profiles through the following equation \citep[see, for instance, ][]{Landi2004}

\begin{equation}
\phi =\frac{1}{2} \arctan \left(\frac{U}{Q}\right).
\end{equation}
The wavelength position used corresponds to the spectral point with maximum $Q(\lambda)$ or $U(\lambda)$ signals. The results for the azimuth are displayed in the top row of Figure~\ref{Azi} over the spatial distribution of the maximum total linear polarisation ($LP=\sqrt{Q_{max}^2+U_{max}^2}$) for three spectral lines from the 850~nm window. To simplify the analysis, we picked a photospheric line sensitive to the magnetic field at lower layers (Fe~{\sc i} 8468~\AA), and the two infrared Ca~{\sc ii} lines that are most sensitive to the atmospheric parameters at slightly different heights in the chromosphere. In this regard, headless (we do not solve the 180 degrees ambiguity) arrows show the inferred average magnetic field azimuth weighted with the strength of the linear polarisation signals. Starting with the photosphere, we can detect the presence of twist around the jet with lines that circle it, see (2.2,4.0)~Mm.  This type of pattern can also be recognised at higher layers in the atmosphere indicating that the magnetic field configuration described in \cite{Iijima2017}  leaves a characteristic imprint in the linear polarisation profiles above the added noise signals. 

Regarding the longitudinal component of the magnetic field, we study its spatial distribution by plotting the maximum Stokes $V$ signals (see the bottom row of Fig.~\ref{Azi}). There is a strong concentration at lower heights (leftmost panel) that occupies larger areas as we examine lines that form higher in the atmosphere. Moreover, there is a thin region that corresponds to magnetic field lines that extend beyond the core of the jet (see sky blue lines at the left part of the structure presented in Figure 8 of \cite{Iijima2017}). This thin region is located further from the centre of the jet when we scan higher heights. Finally, linear polarisation signals are situated outside the strong longitudinal field concentrations, in agreement with the spatial distribution of the atmospheric parameters presented in Figure~\ref{Context}.

\section{Inversion of the Stokes profiles}

We perform NLTE inversions of the full Stokes vector synthesised with the {\sc nicole} code after adding a noise signal. In this regard, computing the Stokes profiles, adding noise, and then inverting them, is something that has been done in the past by various authors with photospheric \citep[e.g.][]{OrozcoSuarez2010,Riethmuller2019} and chromospheric \citep[among others,][]{delaCruzRodriguez2012,delaCruzRodriguez2016,Felipe2018,daSilvaSantos2018,Milic2018} lines \citep[see also the reviews of][]{delToroIniesta2016,delaCruzRodriguez2017}. In our case, we aim to complement those mentioned works focusing on the advantages of inverting multiple spectral lines versus a single chromospheric line.

\subsection{Single- versus multi-line inversions}\label{inver1}

We mentioned in previous publications of this series that adding spectral lines that form at different heights and have different sensitivity to the atmospheric parameters helps to understand the solar phenomena. This argument was based on various studies like the spatial distribution of polarisation signals or the response functions to changes in different atmospheric parameters. However, we have not studied so far the benefits of inverting the full spectrum at 850~nm versus, for instance, only fitting the Ca~{\sc ii} 8542~\AA \ line. In this regard, the target of this section is to invert the Stokes profiles for those two cases. 

We use the field of view enclosed by the square in Figure~\ref{fov} and the inversion set up is the same for the two spectral configurations. In particular, and to simplify the test, we use the default inversion configuration available in the {\sc nicole} distribution. In both cases, we initialise the inversion using the HSRA \citep{Gingerich1971} atmosphere as a guess model with a constant magnetic field of 200~G, 30~degrees of inclination and azimuth, and a null microturbulence velocity. The atmospheric model contains 41 points in height with a constant optical depth step of $\Delta \log \tau=0.2$ and covers the height range from  $\log \tau=[-7,1]$. To invert each profile, we perform one single cycle, with a given number of nodes where the response functions \citep{Landi1977} are computed. In particular, we use 6, 4, 3, and 2 nodes for temperature, LOS velocity, the longitudinal, and the horizontal components of the magnetic field, respectively. We set the ``inversions'' parameter to 3, which means that the code randomly perturbs the initial atmosphere at the nodes given for each atmospheric parameter and performs the inversion. Then, it repeats the process three times, and finally picks the results that produced the best value for the goodness of the fit.

\begin{figure*}
\begin{center} 
 \includegraphics[trim=0 0 0 0,width=13.0cm]{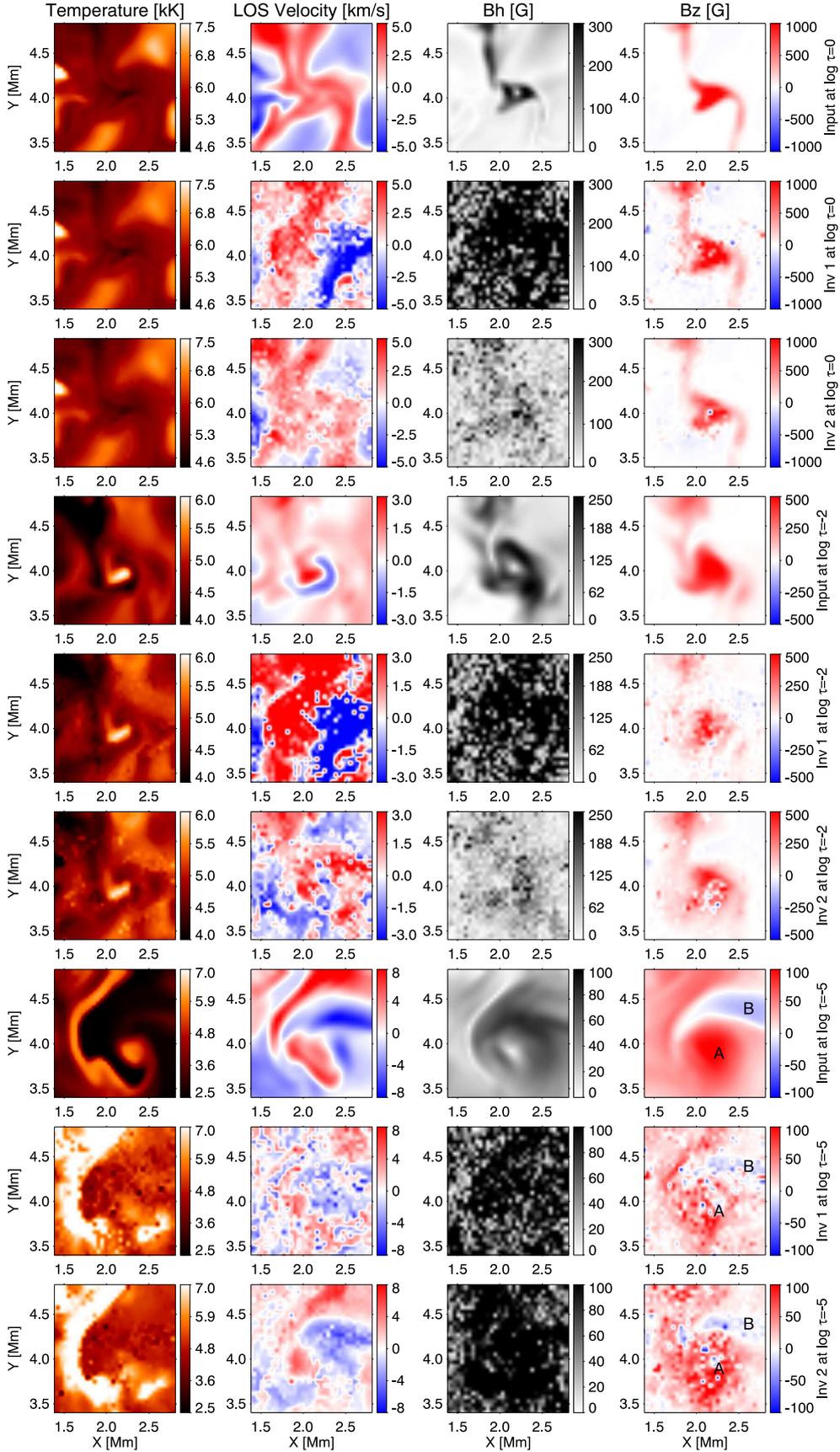}
 \vspace{-0.1cm}
 \caption{Comparison between the input atmospheric parameters, the inferred parameters from the inversion of Ca~{\sc ii}~8542~\AA, and those obtained from the inversion of the full 850~nm window. We plot one after the other, from top to bottom, at $\log \tau$=[0,-2,-5], respectively.}
 \label{inver}
 \end{center}
\end{figure*}

We present in Figure~\ref{inver} the results of the two cases, i.e. single- and multi-line inversions. We chose to display the results at three selected atmospheric layers that are the same ones used in Fig.~\ref{Context}. Each column represents a different physical parameter while rows, from top to bottom, show the input atmosphere, the inferred atmosphere from the fit of the Ca~{\sc ii} 8542~\AA \ spectral line, and that obtained when fitting the complete spectral window. 

Starting with the temperature, both inversion results match the input atmosphere at the lower and middle layers. In the case of upper layers they are similar between inversion configurations although less accurate, i.e. the hot thread on the left side is partially reproduced, but the cold area in the right part of the field of view is not recovered. These results indicate that the Ca~{\sc ii} 8542~\AA \ spectral line alone shows high sensitivity to the temperature from the bottom of the photosphere to the lower-middle chromosphere \citep[see for instance the response functions (RF) presented in][]{QuinteroNoda2016} and is the dominant contribution in the inversion process.  

Examining the LOS velocity results, we start to see differences between the two inversion configurations. There is systematically an improvement at lower layers when inverting additional photospheric lines. For instance, the granulation velocity pattern at $\log\tau=0$, as well as, the core of the jet with a downflow surrounded by material moving upwards (see (2.2,4)~Mm) at $\log\tau=-2$. In the case of upper layers, the multi-line inversion continues increasing the accuracy of the results, i.e. we recover the downflow at (2.2,4)~Mm and the upflowing region that extends towards the right side at (2.5,4.6)~Mm. However, in both cases, some areas are poorly fitted, mainly at $\log\tau=-2$, where even the velocity sign is wrong in most of the locations outside the jet.

Concerning the horizontal component of the magnetic field, we obtain wrong values for the inversions of the Ca~{\sc ii} 8542~\AA \ line everywhere. The erroneous results at lower layers are expected because that spectral line shows low sensitivity there. In this regard, looking at the results for the 850~nm window at lower layers we found an improvement, even in weakly magnetised areas. In the case of upper layers, we do not understand the the reason for the inaccuracy because the Ca~{\sc ii} 8542~\AA \ line shows sensitivity to all the atmospheric parameters at those heights. However, as the results for the 850~nm window at upper layers are also inaccurate, we assume that the large deviations from the input atmosphere are caused by a non-optimal inversion configuration (we focus on this on the following sections).

Finally, regarding the longitudinal component of the magnetic field, the inferred atmosphere is closer to the input parameters for both inversions. However, we see a better match for the case of the multi-line inversions, particularly at lower heights.

\subsection{Improving the inversion configuration}

In the previous section, we found that inverting multiple lines shows an improvement on the inferred atmospheric parameters, mainly at lower heights. However, we also mentioned that the results were sometimes incorrect, even for the multi-line inversions. We stated that the reason could be a non-optimised inversion configuration. Thus, in this section, we aim to improve it. We focus mainly on two elements. On the one hand, the weight the Stokes parameters, as well as the weight different spectral regions of the spectrum, have on the computation of the goodness of the fit ($\chi^2$). On the other hand, the number of nodes used in the inversion process. 

Starting with the first part, we write below a general definition \citep[more specific expressions can be found in the review of][]{delToroIniesta2016} of the goodness of the fit to explain the role of the weight on its computation,

\begin{equation}
\chi^2=\sum_{k=1}^{4}\sum_{i=1}^{M}\left[ I_{k}^{obs}(\lambda_{i})- I_{k}^{syn}(\lambda_{i})\right]^2\times w_{k}(\lambda_{i}),
\end{equation}
where the index $i$ stands for the wavelength points while $k$ defines the four components of the Stokes vector. The labels ``obs" and ``syn" refer to the observed and synthetic data, respectively. Finally, $w_{k}(\lambda_{i})$ is the mentioned weight that is wavelength and Stokes parameter dependent. 

Since early works on inversions assuming LTE conditions, for instance, \cite{RuizCobo1992}, it is customary to use weights that are Stokes parameter dependent. The reason is that the signals of the polarisation profiles are always lower than that of the intensity. Thus, when computing the $\chi^2$, a relative error of the same amplitude in Stokes $I$ and $V$, will have a weaker impact in the latter case as its amplitude is lower. Later on, having in mind the scenarios where some part of the spectrum needs to be removed, e.g. telluric lines in ground-based observations, those weights became wavelength dependent too (see {\sc nicole}'s manual). Moreover, when performing inversions of multiple spectral bands, one sometimes needs to weight more one region of interest than others. This was done, for instance, in the recent work of \cite{daSilvaSantos2018} where the continuum wavelengths observed with ALMA \citep{Wootten2009} were of particular importance. 

In our case, we face the task of inverting together multiple photospheric and chromospheric lines, where wavelength dependent weights could play an important role. For instance, if we have two photospheric lines that we aim to invert simultaneously, although their intensity signals are equivalent, the polarisation signals they produce can display different amplitude. This can be due to different sensitivity to the magnetic field or the spectral properties of each line. Still, the deviations on the amplitude of those polarimetric signals for traditional lines are in the order of a factor two or smaller, e.g. the Fe~{\sc i}~630~nm or the Fe~{\sc i}~525~nm line pairs. However, in the case of chromospheric lines, the differences in polarisation signals respect to those generated by photospheric lines could be larger than a factor 10 (see the bar scale of the different panels of Figure~\ref{Azi}). That large amplitude disparity inevitably reduces the impact they have on the computation of the $\chi^2$. That is, an error in the fit of the Ca~{\sc ii}~8542~\AA \ polarisation signals will be intrinsically much lower than that of the Fe~{\sc i}~8468~\AA \ ($g_{eff}$=2.5) signals when setting the same weight for both spectral lines. Therefore, when inverting two extreme cases like those represented by the two mentioned spectral lines one must be sure that their impact on the computation of the goodness of the fit is properly balanced.

\begin{figure}
\begin{center} 
 \includegraphics[trim=0 0 0 0,width=7.5cm]{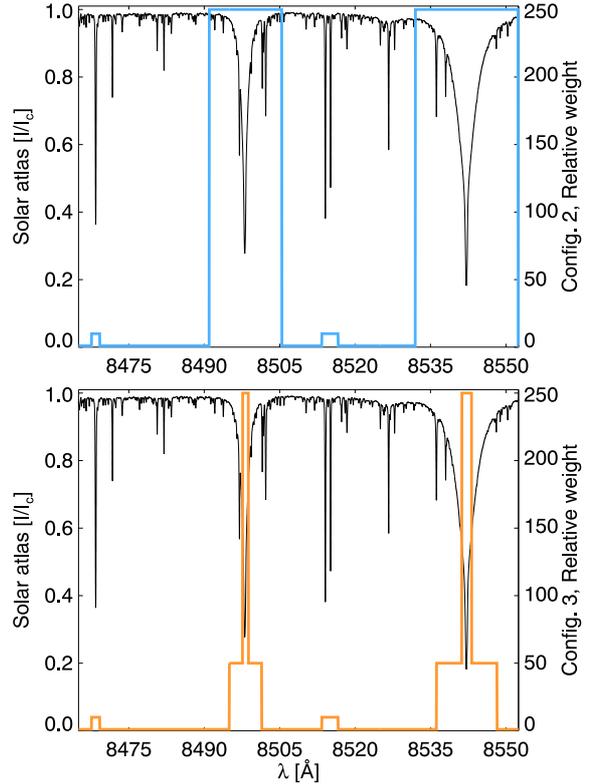}
 \vspace{-0.2cm}
 \caption{Wavelength dependent weights used for the inversion of the Stokes profiles. We use three different configurations, i.e. same weight for all the spectrum (not included here), larger weight for chromospheric lines (top), and more weight for the line core wavelengths of chromospheric lines than the wings (bottom). In the two cases presented here, the photospheric Fe~{\sc i} lines  at 8468 and 8514~\AA \ are weighted more than the rest of the spectrum.}
 \label{weights}
 \end{center}
\end{figure}

\begin{figure*}
\begin{center} 
 \includegraphics[trim=0 0 0 0,width=13.5cm]{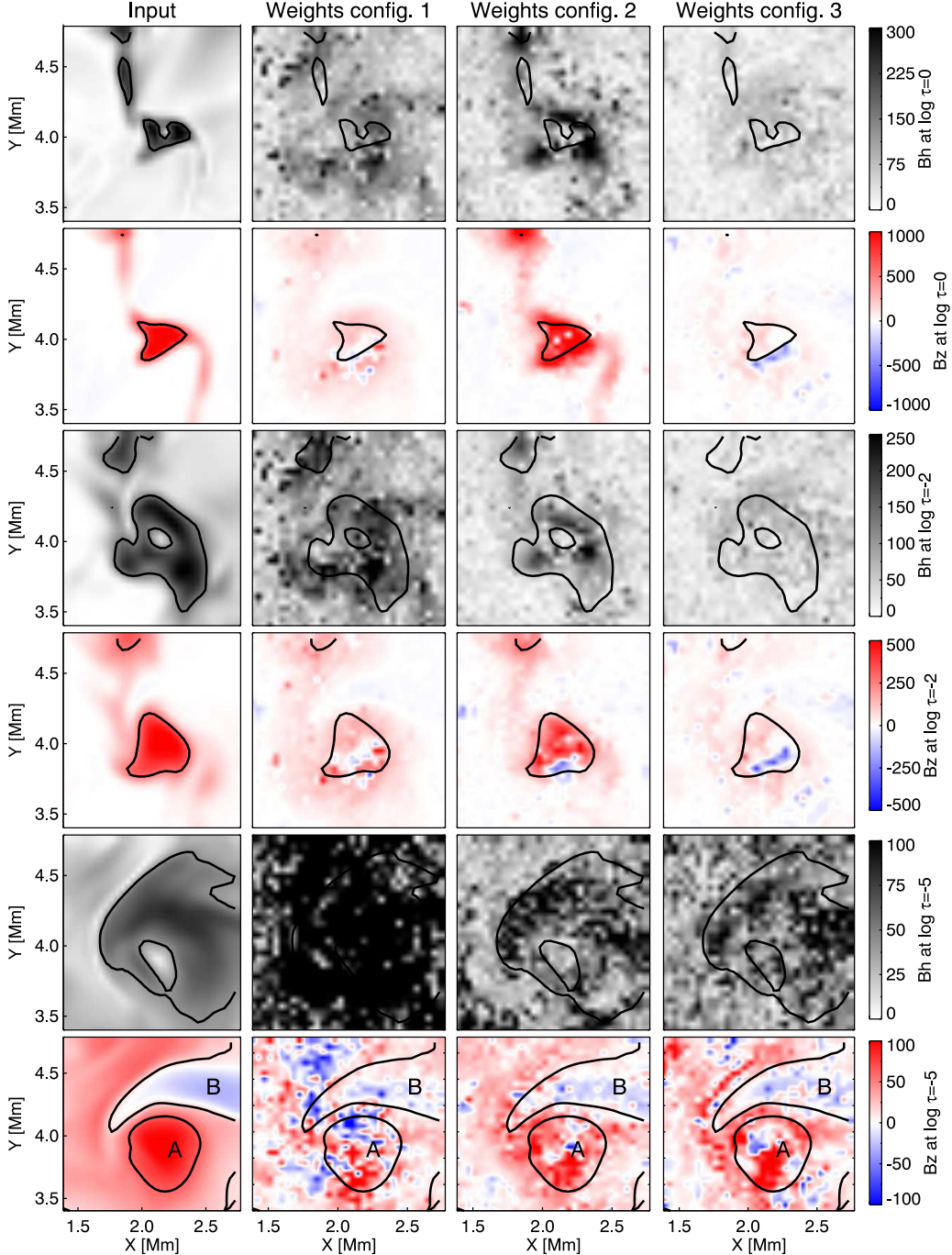}
 \vspace{-0.2cm}
 \caption{Inversion results for different weight configurations. From left to right, input atmosphere and the inferred physical parameters using the weights setting 1, 2, and 3, respectively. From top to bottom, the horizontal and vertical component of the magnetic field at $\log$~$\tau=[0,-2,-5]$. Contour lines designate the spatial distribution of the input atmospheric parameters.}
 \label{weights_compar}
 \end{center}
\end{figure*}

\begin{figure*}
\begin{center} 
 \includegraphics[trim=0 0 0 0,width=15.0cm]{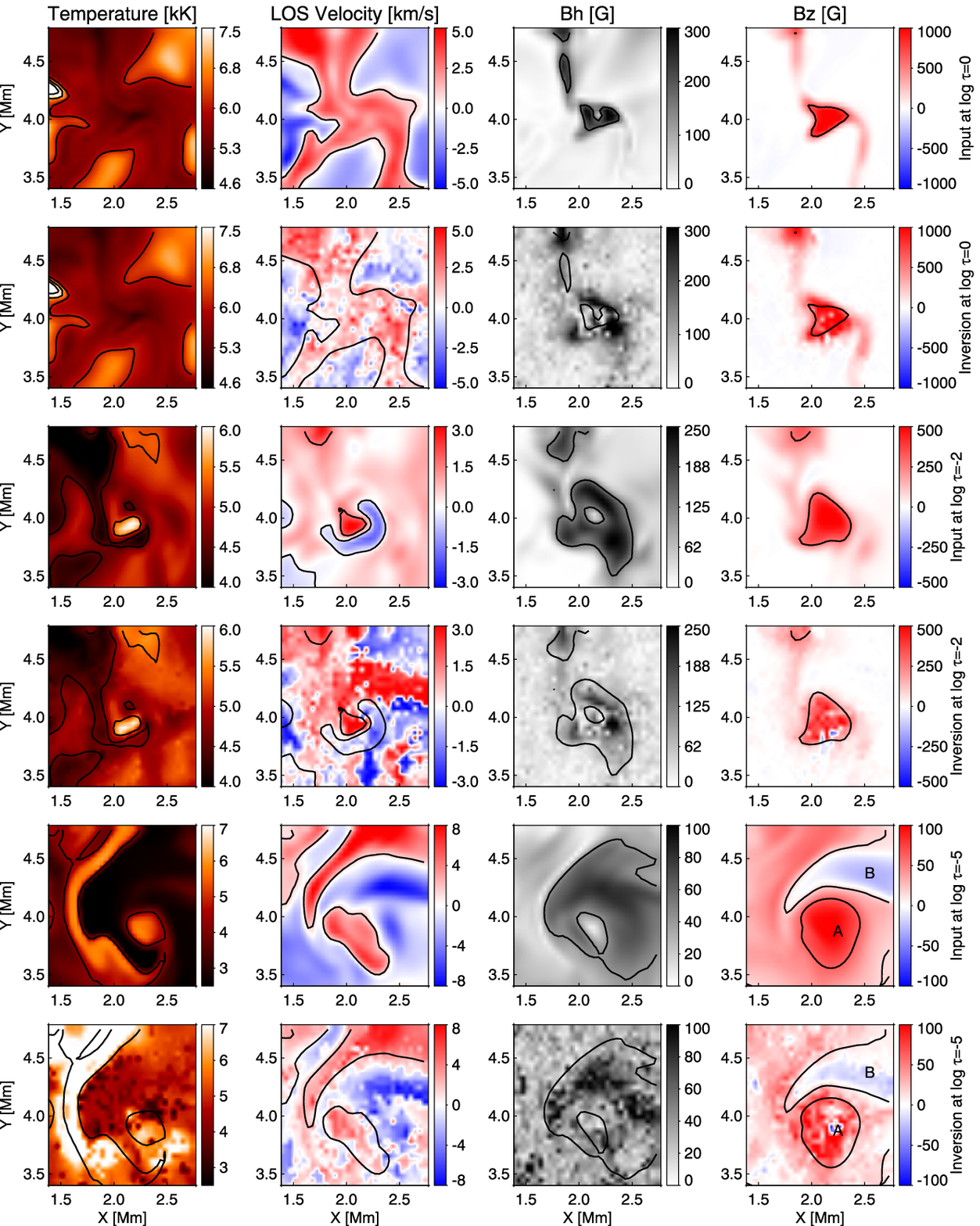}
 \vspace{-0.2cm}
 \caption{Comparison between the input atmospheric parameters and those obtained from the inversion of the spectral lines at the 850~nm window using the weight configuration number 2. We plot one after the other, from top to bottom, at $\log$~$\tau=[0,-2,-5]$, respectively. Contour lines designate the spatial distribution of the input atmospheric parameters.}
 \label{inver850}
 \end{center}
\end{figure*}

In this section, we perform a study to verify the importance of the wavelength dependent weights on the inversion results. We choose different sets of weights, and we run the inversion process with the same configuration used in the previous section. The target is to estimate what could be a reasonable configuration for future instruments that observe the 850~nm window or similar multi-line observations that combine photospheric and chromospheric lines.  In this regard, we use three sets of wavelength-dependent weights. In all cases, the weight ratio among Stokes parameters is the same where we put more weight on the Stokes parameters that show weaker polarisation signals, i.e. (1, 50, 50, 10) for ($I$, $Q$, $U$, $V$), respectively. The first wavelength dependent weight configuration is the default of the code, i.e. identical weight for all the wavelengths (the same that we used in the previous section). The second and third weight configurations we use are presented in Figure~\ref{weights}. The former provides larger values for the entire spectra covered by the chromospheric lines (250 times larger than the continuum), including the wings and the photospheric lines blended with them. The latter configuration gives 250 times more weight to the line core wavelengths and 50 times more weight to the spectral wings than the continuum. Moreover, in both cases, we weight 10 times more the Fe~{\sc i} lines at 8468 and 8514~\AA because they are the most sensitive ones to the magnetic field at the photosphere.

\begin{figure*}
\begin{center} 
 \includegraphics[trim=0 0 0 0,width=17.0cm]{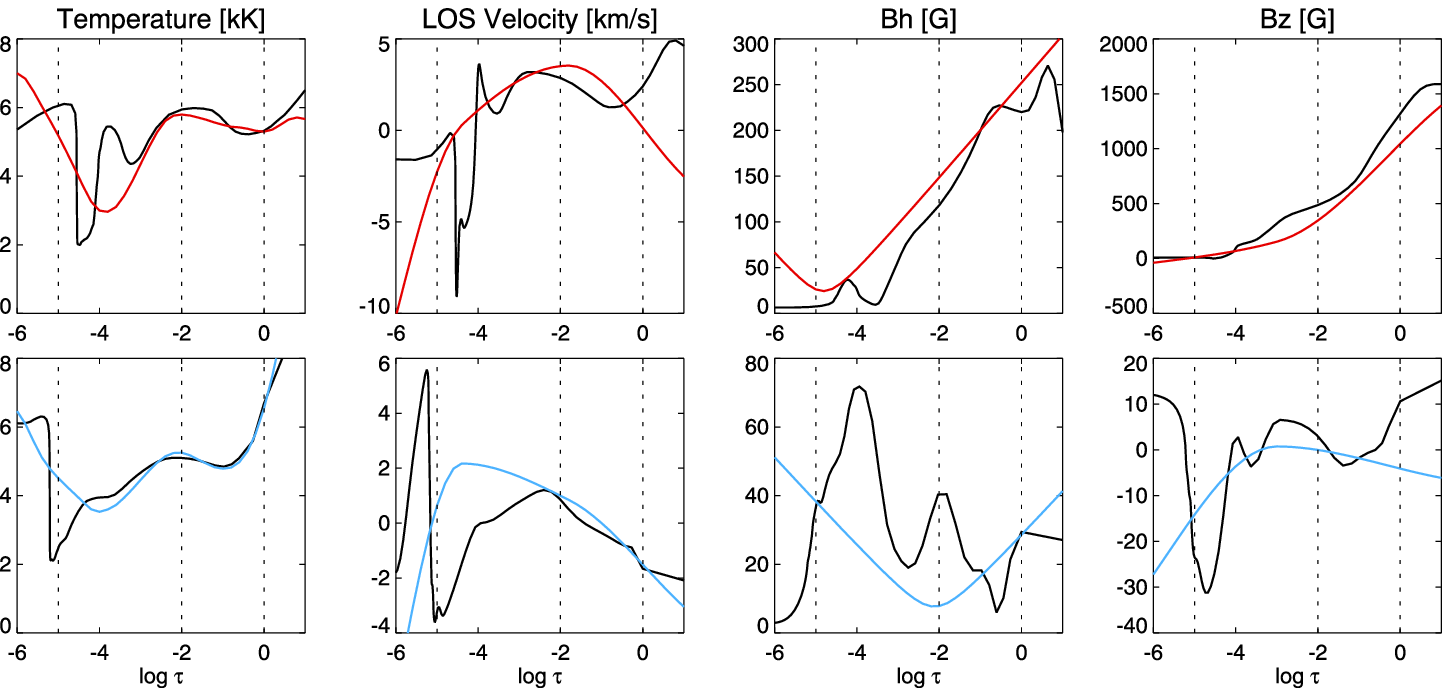}
 \vspace{-0.2cm}
 \caption{Comparison between the input (black) and the inferred (colour) atmosphere for two selected pixels. From left to right, we show the temperature, LOS velocity, the horizontal and vertical component of the magnetic field. Top row corresponds to a pixel close to the ``A'' labelled region while the bottom pixel belongs to the area labelled as ``B'' in Figure~\ref{Context}. Dotted lines designate the heights used in the comparison presented in Figure~\ref{inver850}, i.e. $\log$~$\tau=[0,-2,-5]$.}
 \label{atmoscompare}
 \end{center}
\end{figure*}

We present the results of this comparison in Figure~\ref{weights_compar}. We focus only on the horizontal and vertical component of the magnetic field at different heights to simplify the visualisation. We can see that the first configuration produces incorrect results in general for both magnetic field components (second column from the left). In the case of the second configuration (third column from the left), there is an improvement in all layers and for both components of the magnetic field. Finally, with the third configuration (rightmost column) we obtain accurate results in higher layers although lower heights show a worse match than that obtained with the second configuration of weights. Moreover, this same behaviour is reproduced by the LOS velocity results (not shown here) where the second configuration is the one that yields better results at all atmospheric layers. Therefore, we pick the second configuration to test our capabilities to extract the physical information of the simulated atmosphere from noisy polarimetric spectra. 

We mentioned at the beginning of this section that we also study the impact the number of nodes has on the inversion results. In this regard, starting with the configuration of weights number 2 we examined different combinations of nodes for different atmospheric parameters. However, we did not find a noticeable improvement from the results obtained with the current setup. Therefore, we opt to use the current one as best option to study our capabilities for understanding the physics of the present simulation, and we leave that comparison results outside to reduce the content of this work. 

We repeat the inversion done with the second weight configuration and the same number of nodes but increasing the number of random inversions up to 10. The aim is to slightly reduce the characteristic salt-and-pepper pattern that sometimes appears on the inversion results (see in Figure \ref{weights_compar}, for instance, the inferred longitudinal field at $\log$~$\tau=-5$). However, in the future, it could be useful to also apply some regularisation filter as that implemented in {\sc stic} \citep{delaCruzRodriguez2018}. 

Figure~\ref{inver850} shows the comparison between the input atmosphere and that inferred with {\sc nicole} using the ``optimised'' inversion configuration. We can see again that the temperature results at lower layers are very similar to the original ones. At higher atmospheric layers, we still have some discrepancies in the coldest areas, although the code matches some of them this time. We are not entirely sure about the reasons for these differences, but we try to explain them in the discussion section. Regarding the LOS velocities, the spatial distribution resembles (better than in the first results shown in Figure~\ref{inver}) that of the input atmosphere at all heights. However, we are not able to achieve a good accuracy at $\log$~$\tau=-2$, mainly outside the core of the chromospheric jet, something that we do not understand either. We guess that a computation of the RF in those areas could reveal more information, but we leave that for future studies. In the case of the horizontal component of the magnetic field, we also detect an improvement with respect to previous inversions and a similar spatial distribution to the input one. Finally, concerning the longitudinal magnetic field, we have excellent results, even at the highest layers where the weak, in order of 50~G, opposite polarity field (label~B) is well reproduced as well as the core of the jet (label~A).

\subsection{How accurate are the results?}\label{accuracy}

Based on the results presented in Figure~\ref{inver850} we can say that the inferred parameters match in general the input atmosphere at the selected heights. Those heights were chosen based on previous studies on the RF to changes in the physical parameters for the lines of interest using the FALC atmosphere. In particular, the continuum and ``weak'', in terms of low-intensity depression, spectral lines show a peak at bottom layers close to $\log$~$\tau=0$. The core of stronger photospheric lines shows the same peak in the RF at around  $\log$~$\tau=-2$, while the core of the Ca~{\sc ii}~8542~\AA \ line displays a maximum centred around $\log$~$\tau=-5$. It is important to bear in mind that, as the RF depend on the atmospheric model used, the mentioned results do not necessarily satisfy the present case. However, we assumed that the said layers could be a good reference point for this work.  This raises the question of whether we can reproduce the small scale variations of the simulated atmosphere as well. In other words, if we choose a different set of optical depths, would we find a good correlation too? 

The answer to this question is no, and for illustrating purposes, we plot in Figure~\ref{atmoscompare} the vertical stratification of the atmospheric parameters for two selected pixels. Those pixels are located close to the areas labelled as A and B, respectively, and they show good results for the longitudinal field at $\log$~$\tau=-5$. If we examine the inferred atmospheres (colour), we find that the inversion is not able to reproduce the steep variations that take place in the simulation at short height scales. However, it follows the tendency concerning gradients, absolute value and sign of the atmospheric parameters. Most importantly, when examining the regions highlighted with dotted lines (the same heights used in Figure~\ref{inver850}), we can see that there is where the code usually gets closer to the original atmosphere. Hence the reason why we obtained a good agreement in previous sections. We believe that the cause for this behaviour is that the RF of the whole set of spectral lines is large at those optical depths. In other words, when analysing future observations of this set of spectral lines (or a similar combination), we can represent the results at discrete layers like the ones used here. Also, we could even believe the general properties, concerning gradients, absolute values, and the sign of the atmospheric parameters at heights in between those layers.

\section{Discussion}

We ascertained that the coldest regions in the upper layers are systematically hotter in the inversion results. If we revisit the work of \cite{delaCruzRodriguez2012}, we can find a similar problem where the inferred temperatures are consistently lower at higher layers. They explained that this is because the synthesis was done assuming 3D with {\sc multi} \citep{Carlsson1986,Leenaarts2009MULTI} while the inversion was performed in 1D with {\sc nicole}. However, in our case, we cannot use the same argument as we did both the forward modelling and the inversion with {\sc nicole}. Thus, we should look for additional explanations. 

\begin{figure}
\begin{center} 
 \includegraphics[trim=0 0 0 0,width=6.5cm]{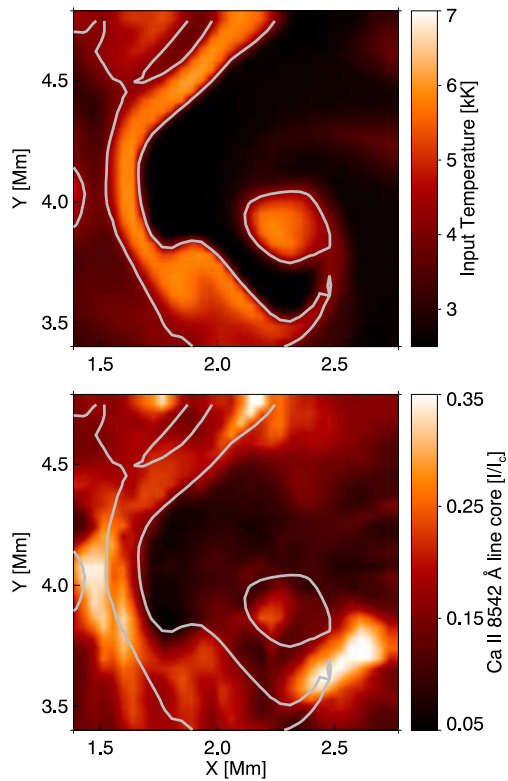}
 \vspace{-0.2cm}
 \caption{Comparison between the input temperature from the simulation (top) and the line core intensity for the Ca~{\sc ii} 8542~\AA \ chromospheric line (bottom). Contours display the spatial distribution of the input atmosphere.}
 \label{linecore}
 \end{center}
\end{figure}

One option could be that, as the temperature in the simulation at chromospheric layers is low, the Ca~{\sc ii} lines, the ones that form at those heights, are not sensitive to the small temperature variations we see in the plots. To verify this assumption, we show in Figure~\ref{linecore} the temperature at $\log$~$\tau=-5.0$ and the Ca~{\sc ii}~8542~\AA \ line core intensity.

\begin{figure}
\begin{center} 
 \includegraphics[trim=0 0 0 0,width=6.3cm]{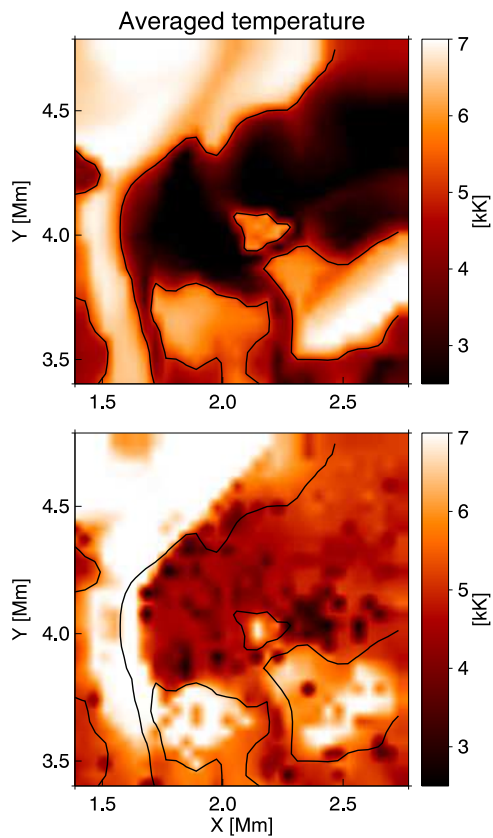}
 \vspace{-0.2cm}
 \caption{Comparison between the input atmosphere (top) and the one inferred from the inversions (bottom) when averaging over the optical depth range comprehended by $\log$~$\tau=[-5.2,-4.8]$. Contours represent the averaged input atmosphere's features.}
 \label{averaged}
 \end{center}
\end{figure}

Examining the results, we find intensity variations, from 0.05 to 0.15 of $I_c$, at around (2,4)~Mm that indeed resemble the input atmosphere. At the same time, there are enhanced patches at (1.5,4) and (2.5,3.7)~Mm that do not correspond to any hot feature in the simulation. Moreover, those enhanced areas are reproduced in the recovered temperature stratification (see Figure~\ref{inver850}). Therefore, this leads us to a different possibility; the line core intensity and consequently the inferred temperature that fits it, could correspond to a given range of optical depth layers instead of a single height. Thus, if we average the temperature, for instance between $\log$~$\tau=[-5.2,-4.8]$ we should find a similar pattern between the averaged input and inferred atmospheres. We display this comparison in Figure~\ref{averaged} where we can see that indeed, the temperature spatial distribution is similar, both showing the enhancements found in the line core intensity. It is true that the central part still shows cool areas, something that it is barely perceptible in the inversion results, so we should continue investigating the reason of the lack of accuracy for the inferred temperature at higher layers. One option that we want to test in the future is to manually increase the density of nodes, for instance, between $\log$~$\tau=[-5.5,-4.5]$. This could help to reproduce the stark temperature jumps that can be seen in Figure~\ref{atmoscompare}. More important is that, in spite of the improvement for the temperature comparisons, if we average the LOS velocity or magnetic field components over the same optical depth range, the results are worse, i.e. we see fewer similarities than when selecting the three specific layers used in previous sections. For this reason, and taking into account the explanations given in Section~\ref{accuracy}, we still believe that it is better to plot the spatial distribution of atmospheric parameters at selected optical depth layers where the RF have a significant value.

\section{Summary}

We examined the properties of the simulation presented in \cite{Iijima2017} that contains a chromospheric jet that, in its emerging phase, extends several Mm above the solar surface. At the low atmosphere, the temperature at the core of the structure is higher than its surroundings, and it is dominated by downward velocities, while outer areas are colder and display upward motions. Regarding the magnetic field, the core of the structure shows a strong longitudinal component of positive sign encompassed by horizontal fields and a weaker longitudinal field of opposite polarity.

The next step we took was to examine the synthetic Stokes profiles. The amplitude of linear polarisation signals for chromospheric lines is around $1\times10^{-3}$ of $I_c$ while ten times more significant for the case of circular polarisation signals. The latter means that with average integration times, assuming diffraction limit, modern telescopes should be able to achieve those signals. The twist of the jet governs the magnetic field azimuth derived from the linear polarisation signals with field lines that move away from the core of the structure, arching and bending around it. In the case of the circular polarisation signals, they are strong in the central part of the magnetic feature and show rounded shapes similar to that of the simulated longitudinal magnetic component. 

We took a step further, and we employed the inversion code {\sc nicole} aiming to comprehend the diagnostic potential of the 850~nm window for inferring the atmospheric information of the chromospheric jet through NLTE inversions. We started with a simple study to corroborate that the inversion of multiple spectral lines with different height of formation provides better results than when fitting a single spectral line. We, thus, inverted the synthetic Ca~{\sc ii}~8542~\AA \ signals and compared the results to those obtained when including in the inversion all the spectral lines of the 850~nm window. Interestingly, we found no noticeable improvement on the temperature stratification, which means that the Ca~{\sc ii}~8542~\AA \  line represents the dominant contribution for that physical parameter. For the rest of the atmospheric parameters, we indeed found improvement when including the photospheric lines. However, this mainly happens at lower layers. Moreover, we discovered that the results at upper layers were incorrect in both cases what led us to a new set of tests to find the cause of this problem.

We believe that the main reason is the importance the different spectral lines have in the computation of the goodness of the fit. When using the same weight for photospheric and chromospheric lines on the calculation of $\chi^2$, the latter have an impact in the order of 10 or 100 times lower than the former for the polarisation Stokes parameters. Also, this is independent of where the lines form, and where their maximum sensitivity lies. Therefore, we tested this hypothesis with different wavelength dependent weight configurations where chromospheric lines were weighted more than the rest of spectral lines. The results revealed that there is an improvement in all the inferred atmospheric parameters leading to a better resemblance to the input atmosphere. 

We conclude then that multi-line inversions benefit from the predicted increase of sensitivity and height coverage of the RF mentioned in previous works of this series. But the user should be aware that the computation of the $\chi^2$ requires additional fine-tuning not needed when performing single line inversions. Hence the possibility of choosing wavelength dependent weights in the modern NLTE codes like {\sc nicole}, {\sc snapi} or {\sc stic}. Besides that, we can close this work saying that the Ca~{\sc ii}~8542~\AA \ remains as one of the most promising lines for understanding the magnetism of the chromosphere, and if we complement it with additional photospheric lines we could seamlessly constrain the atmospheric parameters in the low atmosphere.

\section*{Acknowledgements}

C.~Quintero Noda acknowledges the support of the ISAS/JAXA International Top Young Fellowship (ITYF) and the JSPS KAKENHI Grant Number 18K13596. H.~Iijima is supported by MEXT/JSPS KAKENHI Grant Number JP15H05816. This work was supported by the funding for the international collaboration mission (SUNRISE-3) of ISAS/JAXA and the JSPS KAKENHI Grant Number 18H03723 and 18H05234, and by the Research Council of Norway through its Centres of Excellence scheme, project number 262622. J.~de la Cruz Rodr\'{i}guez is supported by grants from the Swedish Research Council (2015-03994), the Swedish National Space Board (128/15) and the Swedish Civil Contingencies Agency (MSB). This project has received funding from the European Research Council (ERC) under the European Union's Horizon 2020 research and innovation programme (SUNMAG, grant agreement 759548). This work has also been supported by Spanish Ministry of Economy and Competitiveness through the project ESP-2016-77548-C5-1-R. D.~Orozco Su\'{a}rez also acknowledges financial support through the Ram\'{o}n y Cajal fellowships.

\bibliographystyle{mnras} 
\bibliography{multiline3} 

\bsp	
\label{lastpage}
\end{document}